\documentclass{iau}

\usepackage{amsmath}
\usepackage{graphicx}
\usepackage{multirow}

\renewcommand{\vec}[1]{\mbox{\boldmath $#1$}}

\begin{document}

\lefttitle{L.\,L.~Kitchatinov}
\righttitle{Near-surface shear layer of solar rotation}

\jnlPage{1}{4}
\jnlDoiYr{2024}
\doival{10.1017/xxxxx}

\aopheadtitle{Proceedings IAU Symposium }
\editors{A.\,V.~Getling \&  L.\,L.~Kitchatinov, eds.}

\title{Near-surface shear layer of solar rotation:\\ origin and significance}

\author{Leonid Kitchatinov$^{1,2}$}
\affiliation{$^1$Institute of Solar-Terrestrial Physics, Lermontov Str. 126A, 664033, Irkutsk, Russia;
\email{kit@iszf.irk.ru}}
\affiliation{$^2$Pulkovo Astronomical Observatory, Pulkovskoe Sh. 65, 196140, St-Petersburg, Russia}

\begin{abstract}
Helioseismology has discovered a thin layer beneath the solar surface where the rotation rate increases rapidly with depth. The normalized rotational shear in the upper 10 Mm of the layer is constant with latitude. Differential rotation theory explains such a rotational state by a radial-type anisotropy of the near-surface convection and a short correlation time of convective turbulence compared to the rotation period. The shear layer is the main driver of the global meridional circulation.
\end{abstract}

\begin{keywords}
Sun: interior, Sun: rotation, turbulence
\end{keywords}

\maketitle

\section{Introduction}
Helioseismology has detected a steep increase in the rotation rate with depth just below  the surface \citep{Thompson_EA_96,Schou_EA_98Helioseismic_DR}. The subsurface radial shear exceeds the latitudinal shear seen on the solar surface. The potential importance of the large differential rotation in the near-surface shear layer (NSSL) for the solar dynamo \citep{Brandenburg_05_NSSL_dynamo,Pipin_Kosovichev_11_NSSL_dynamo} has triggered a discussion on the origin of the layer \citep[see e.g.][]{Kit_13_IAUS294,Hotta_EA_15,GB_19,JC_21}. The discussion has not led to a consensus yet.

It is noteworthy that although the near-surface shear varies with latitude, it varies coherently with the rotation rate, so that the normalised shear
\begin{equation}
    \frac{r}{\Omega}\frac{\partial\Omega}{\partial r} \simeq -1
    \label{1}
\end{equation}
in the upper 10 Mm of the solar convection zone is essentially constant \citep{Barekat_EA_14}. An adequate theory should reproduce this remarkable property.

The rotational shear proportional to the rotation rate is a robust prediction of  differential rotation theory for the case of a short correlation time of the convective turbulence compared to the rotation period \citep{Ruediger_89}. The upper 10 Mm of the NSSL belong to this case. The constant normalised shear then follows from the standard stress-free condition for the upper boundary.
\section{Differential rotation theory for NSSL}
Thermal convection is driven by buoyancy forces pointing up or down the radius. Convective turbulence can therefore be anisotropic, with different intensities of radial and horizontal mixing. \citet{Lebedinsky_41} was probably the first to point out that the influence of anisotropic turbulence on global rotation is not limited to the smoothing of rotational inhomogeneity by turbulent viscosity. Anisotropy allows the turbulence to generate differential rotation through non-diffusive fluxes of angular momentum \citep[see][for pictorial explanation]{Kit_05_PhyU}. The Lebedinsky effect is now known as the $\Lambda$-effect \citep[][p.37]{Ruediger_89}.

The turbulent transport of momentum is accounted for in the mean field hydrodynamics by the Reynolds stress, $R_{ij} = -\rho\langle u_i u_j\rangle$, where the angular brackets mean averaging, $\rho$ is the density and $\mathbf{u}$ is the turbulent velocity. The differential rotation theory distinguishes two parts in the stress tensor,
\begin{equation}
    R_{ij} = R_{ij}^\nu + R_{ij}^\Lambda,
    \label{2}
\end{equation}
which are responsible for the turbulent viscosity and the $\Lambda$-effect.

\begin{figure}[htb]
    \centering
    \includegraphics[width = 9 truecm]{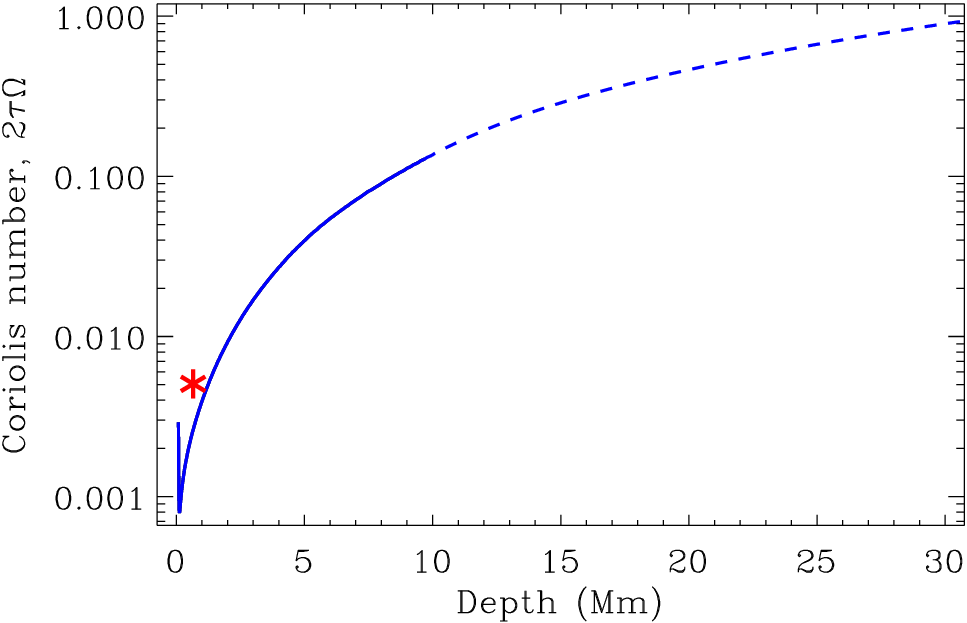}
    \caption{Depth profile of the Coriolis number (\ref{3}) in the NSSL.
             The solid line shows the profile for the depth range where \citet{Barekat_EA_14} found the constant shear of Eq.\,(\ref{1}). The profile for greater depths is shown dashed. The star indicates an estimate for the solar granulation.}
    \label{f1}
\end{figure}

The key parameter of the theory is the dimensionless factor
\begin{equation}
    \Omega^* = 2\tau\Omega
    \label{3}
\end{equation}
of the Coriolis force in the normalised equation of motion; $\tau$ is the convective turnover time. The Coriolis number (\ref{3}) measures the intensity of the interaction between convection and rotation. The main difficulty with the differential rotation theory was a large value of the Coriolis number in the depth of the solar/stellar convection zone. Therefore, the theory has to be non-linear in this number \citep{RKH_13}. However, this difficulty does not apply to the depth range where \citet{Barekat_EA_14} found the constant normalised shear of Eq.\,(\ref{1}). Figure\,\ref{f1} shows small Coriolis number for these depths.

The total depth of the NSSL is conventionally assumed to be about 30\,Mm. The radial shear changes sign around this depth \citep{Schou_EA_98Helioseismic_DR}. The shear value decreases and the relation (\ref{1}) is violated as this depth is approached \citep{Komm_22}. It is a matter of definition whether the depths near 30\,Mm, where the radial shear is no longer large compared to the latitudinal shear, should be assigned to the NSSL. This paper focuses on the upper 10\,Mm of the convection zone where the relation (\ref{1}) constrains possible explanations of the NSSL.

A further complication to the theory is that not only the Reynolds stress but also the meridional flow can also cause differential rotation. However, the meridional flow does not enter the stress-free boundary condition,
\begin{equation}
    R_{r\phi} = R^\Lambda_{r\phi} + R^\nu_{r\phi} = 0\ \ \mathrm{at}\ r = R_\odot,
    \label{4}
\end{equation}
which controls the surface shear \citep{Kit_13_IAUS294}.

The smallness of the Coriolis number and the condition (\ref{4}) simplify matters to such an extent that the constancy of the normalised shear with latitude can be proved without addressing any specific approximation of mean-field theory. Since the dependence on angular velocity enters via the Coriolis number, the Reynolds stress can be linearised in angular velocity. The general structure for the real $R^\Lambda_{ij}$ tensor linear in pseudo-vector $\vec\Omega$ reads
\begin{equation}
    R^\Lambda_{ij} = -\rho\nu_\Lambda\left(\hat{r}_i\varepsilon_{jkl} +
    \hat{r}_j\varepsilon_{ikl}\right)\hat{r}_k\Omega_l ,
    \label{5}
\end{equation}
where $\hat{\mathbf{r}} = {\mathbf{r}}/r$ is the radial unit vector, $\nu_\Lambda$ is a yet indefinite latitude-independent constant, $\varepsilon_{jkl}$ is the fully antisymmetric unit tensor and repetition of subscripts means summation. The viscous part of the Reynolds stress is given by the viscosity tensor ${N}_{ijkl}$
\begin{equation}
    R^\nu_{ij} = \rho {N}_{ijkl}(\nabla_l V_k) ,
    \label{6}
\end{equation}
where $\vec V$ is the large-scale velocity, which includes rotation and the meridional circulation. General expression for the viscosity tensor in the case of horizontally isotropic turbulence ($\langle u^2_\phi\rangle = \langle u^2_\theta \rangle \neq \langle u^2_r\rangle$) is
\begin{eqnarray}
    {N}_{ijkl} &=& \nu_1\left(\delta_{ik}\delta_{jl}
    + \delta_{jk}\delta_{il}\right)
    + \nu_2\delta_{ij}\delta_{kl} + \nu_3\left(\delta_{ik}\hat{r}_j\hat{r}_l
    + \delta_{jk}\hat{r}_i\hat{r}_l\right)
    \nonumber \\
    &+& \nu_4\left(\delta_{il}\hat{r}_j\hat{r}_k + \delta_{jl}\hat{r}_i\hat{r}_k\right)
    + \nu_5\delta_{ij}\hat{r}_k\hat{r}_l + \nu_6\delta_{kl}\hat{r}_i\hat{r}_j
    + \nu_7\hat{r}_i\hat{r}_j\hat{r}_k\hat{r}_l .
    \label{7}
\end{eqnarray}
Substituting Eqs.\,(\ref{5}-\ref{7}) into Eq.\,(\ref{4}) gives the expression for the normalised shear
\begin{equation}
    \frac{r}{\Omega}\frac{\partial\Omega}{\partial r} = - \frac{\nu_\Lambda}{\nu_1+\nu_3}
    \label{8}
\end{equation}
in terms of the eddy transport coefficients for the case of a horizontally isotropic background turbulence and small Coriolis number.

The normalised shear (\ref{8}) is constant with latitude. This constancy is therefore a robust result of mean-field theory independent of any particular theoretical tool used to  derive the shear. However, the value of the surface shear depends on the - necessarily approximate - theoretical tool used for the evaluation. Derivations with the quasi-linear approximation give \citep{Kit_23}
\begin{equation}
    \frac{r}{\Omega}\frac{\partial\Omega}{\partial r} =
    \frac{\langle u_h^2\rangle}{\langle u_r^2\rangle} - 2 ,
    \label{9}
\end{equation}
where $u^2_h = u_\phi^2 + u_\theta^2$ is the intensity of horizontal mixing. Shear is negative for anisotropy of radial type, $\langle u_r^2\rangle > \langle u^2_\phi\rangle = \langle u^2_\theta\rangle$, as it should be \citep{Lebedinsky_41}.

Equation (\ref{9}) reproduces the seismically detected shear (\ref{1}) for $\langle u_h^2\rangle/\langle u_r^2\rangle = 1$. This intensity ratio is in agreement with the numerical experiment on the NSSL by \citet{Kitiashvili_EA_23}. The anisotropy parameter $A_V$ of their fig.\,4 corresponds to $\langle u_r^2\rangle \simeq \langle u_h^2\rangle$ {\em within} the convective zone.
\section{NSSL and meridional flow}
The global meridional flow is an important component of the flux transport dynamo models for solar activity. The importance of the NSSL for the solar dynamo can be mediated by the NSSL relation to the meridional flow. The flow is currently understood to result from a slight imbalance between strong centrifugal and baroclinic drivers of the flow \citep[see review by][for further details]{Hazra_EA_23SSRv}.

\begin{figure}[htb]
    \centering
    \includegraphics[height= 5 truecm]{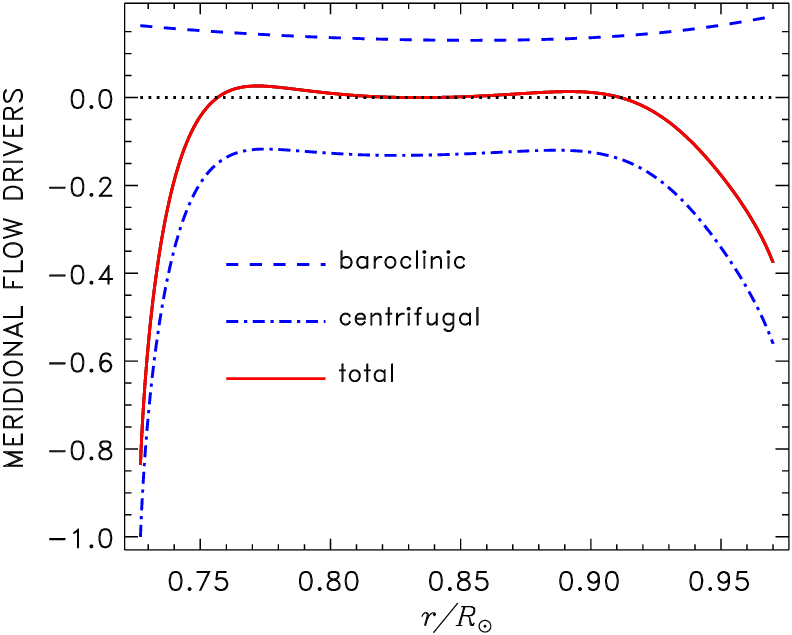}
    \hspace{0.1 truecm}
    \includegraphics[height= 5 truecm]{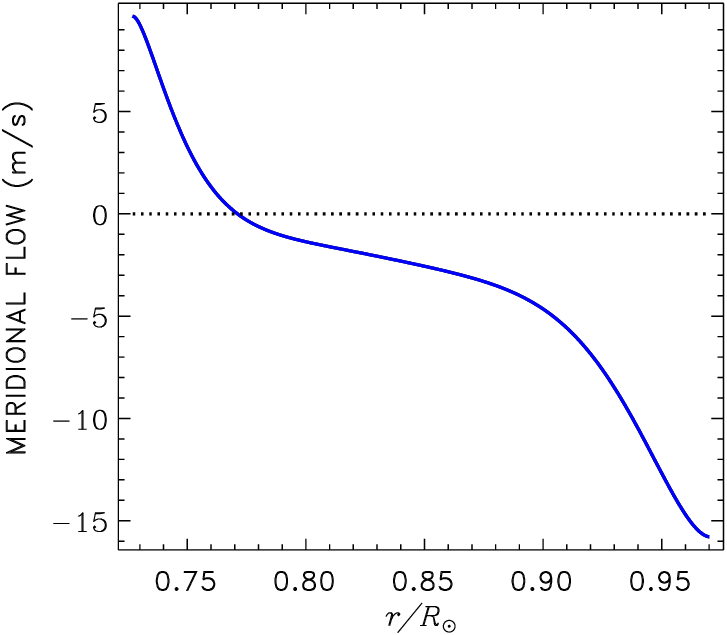}
    \caption{Centrifugal and baroclinic drivers of the meridional flow and their
        sum (left) and the resulting meridional flow (right) as functions of radius for the 45$^\circ$ latitude according to the mean-field model of \citet{Kit_Ole_11DR}.}
    \label{f2}
\end{figure}

The profiles of the two drivers and the resulting meridional flow computed with a mean-field hydrodynamical model are shown in Fig.\,\ref{f2}. The two drivers are almost balanced in the bulk of the convection zone. Close to the boundaries, a complete set of the boundary conditions does not allow the additional condition of the thermo-rotational balance to be satisfied. The increase of the centrifugal driving term near the boundaries violates the balance. Accordingly, the meridional flow of Fig.\,\ref{f2} reaches its maximum velocity at the boundaries and decreases inside the convection zone, in agreement with the seismological detections by \citet{Rajaguru_Antia15} and \citet{Gizon_EA_20Single_cell}. The meridional flow is generated in the NSSL.
\section{Conclusions}
The NSSL is a consequence of the radial type of anisotropy of the near-surface convection, and the condition (\ref{4}) that the external azimuthal force is zero. The constancy of the normalised surface shear (\ref{1}) with latitude is a consequence of the short convective turnover time relative to the rotation period in the depth range where the constancy was found.

Global meridional flow is excited in the boundary layers near the top and bottom of the convection zone. The large rotational shear in the NSSL drives the meridional circulation by its non-conservative centrifugal force.
\acknowledgements{The author acknowledges financial support from the Ministry of Science and High Education of the Russian Federation.}
\bibliographystyle{iaulike}

\end{document}